# Distribution of carrier multiplication rates in CdSe and InAs nanocrystals

Eran Rabani[(a)] and Roi Baer[(b)]
(a) School of Chemistry, The Raymond and Beverly Sackler Faculty of Exact Sciences, Tel Aviv University, Tel Aviv 69978 Israel; (b) Institute of Chemistry and the Fritz Haber Center for Molecular Dynamics, The Hebrew University of Jerusalem, Jerusalem 91904 Israel.

The distribution of rates of carrier multiplication (CM) following photon absorption is calculated for semiconductor nanocrystals (NCs). The NC electronic structure is described using a screened pseudopotential method known to give reliable description of NC excitons. The rates of biexciton generation are calculated using Fermi's golden rule with all relevant Coulomb matrix elements, taking into account proper selection rules. In CdSe and InAs NCs we find a broad distribution biexciton generation rates depending strongly on the exciton energy and size of the NC. The process becomes inefficient for NC exceeding 3 nm in diameter in the photon energy range of 2-3 times the band gap. PACS numbers: 78.67.Bf, 71.35.-y.

Carrier multiplication is a process where several excitons are generated upon the absorption of a single photon in semiconductors [1]. Strict selection rules and competing processes in the bulk allow observation of CM only at energies larger than 5 times the band gap ($E_g$) [2]. In NCs, where quantum confinement effects are important, CM was anticipated at lower photon energies [3]. Indeed, CM in NCs has been reported recently for several systems, such as PbSe and PbS [4-8], PbTe [9], CdSe [6], InAs [10, 11] and Si [12]. These studies showed that the threshold for CM was material dependent, occurring at ~2-3 $E_g$, with an efficiency that was size and band-gap independent [4, 5, 11, 12]. However, several recent studies have questioned the efficiency of CM in NCs, in particular for CdSe [13] and InAs [14]. The goal of the present letter is to address this controversy.

The theory of CM in bulk is based on the concept of impact ionization [15], by which the photon first creates an exciton, composed of the negative electron and positive hole, each having an effective mass depending on the band structure of the crystal. The lighter particle of the pair takes most of the kinetic energy and eventually looses part of this energy by creating additional charge carriers. For NCs, several theoretical approaches for describing CM have been proposed [4, 6, 16-19]. Efros, Nozik and their co-workers[4, 16] developed a time-dependent density matrix formalism taking into account the populations and coherences of single exciton coupled to a single biexciton state. Using this model, in conjunction with an effective mass theory, they developed a theory of impact ionization obtaining expression for the ratio of exciton to biexciton populations at steady states, which depend on the decay rate of the charged particle in the exciton into a trion. This approach treats a single trion neglecting the fact that the charged particle decays into a dense manifold of trions. Under such circumstance a rate approach might be more appropriate, describing the decay as an incoherent impact ionization process [17], similar to the treatment of the process in bulk. This point of view was elaborated by Allan and Delerue [18] using a tight binding model and Franceschetti *et al.* [17] based on a semiempirical pseudopotential density of states calculation. A third approach, assumed a direct generation of a biexciton following absorption of light, suggested by Schaller *et al.* [6].

The previous theoretical work established general conditions for efficient CM, that the frequency (in a coherent model [16]) or rate (in an incoherent model [17]) of bi-exciton generation must be faster than the rate $\gamma_1$ of single exciton decay via other channels. In the coherent theory $\gamma_1 \ll W_C/\hbar$ while in the incoherent case $\gamma_1 \ll 2\pi W_C^2 \rho_{sb}/\hbar$ where $\rho_{sb}$ is the density of biexciton states at the energy of the single exciton [18]. The incoherent result further assumes that $W_C/\hbar \ll \gamma_2$, where $\gamma_2$ is the decay rate of biexciton state to the lowest energy biexciton, and thus the Fermi golden rule applies. These theories predict efficient CM for the aforementioned nanoparticles based on experimental estimates of $\gamma_1 \approx 1\,\text{ps}^{-1}$ and estimating $W_C$ or the impact ionization rate of ~ $10\,\text{ps}^{-1}$. Furthermore, CM sets in at energies below $3E_g$ [4, 17].

For a reliable estimate of CM one requires: (a) A quantitative accurate account of the electronic structure of the NC, especially the highly excited states. (b) A description of the dense manifold of single and bi-excitonic states. (c) A theory of electron correlation that can explain the formation of biexcitons fully consistent with the single particle electronic structure. So far, theoretical treatments of CM in NCs have not met all three requirements at once. A high level work on CM in small clusters was recently submitted by Isborn *et al.* [19] however, the relevance for NCs is not immediately obvious.

In this letter, we present a framework that meets the above requirements. Using an atomistic semiempirical pseudopotential method that captures realistically the density of electronic states, we accurately deduce the density of excitons and biexcitons and calculate the Coulomb matrix elements even at energies high above the band gap. The detailed framework we develop allows us to study the effect of NC size (up to a diameter of ~3 nm and ~2000 electrons), photon energy (up to $3E_g$) and composition (CdSe and InAs NCs) on the process of CM.

We consider CM for two prototype NCs, CdSe (II-VI) and InAs (III-V). The local screened pseudopotentials were fitted to reproduce the experimental bulk band-gap and effective masses for CdSe [20] and InAs [21], neglecting spin orbit coupling [22]. Furthermore, ligand potentials are used to represent the passivation layer [20]. Once the potential is determined, the resulting single-particle Schrödinger equation is solved in real space by the filter-diagonalization (FD) technique [23, 24]. FD allows construction of an eigen-subspace of all energy levels up to $3E_g$ above the Fermi energy. From this, the density of states (DOS) is calculated by energy binning. As a check on the FD we also employed an alternative Monte Carlo method [25] which computes directly the DOS, $\pi^{-1} \operatorname{Im} Tr\left[\left(E - H + i\gamma\right)^{-1}\right]$. Using binning or self convolutions of the DOS, the exciton (DOSX) and bi-exciton (DOSXX) density of states can be determined.

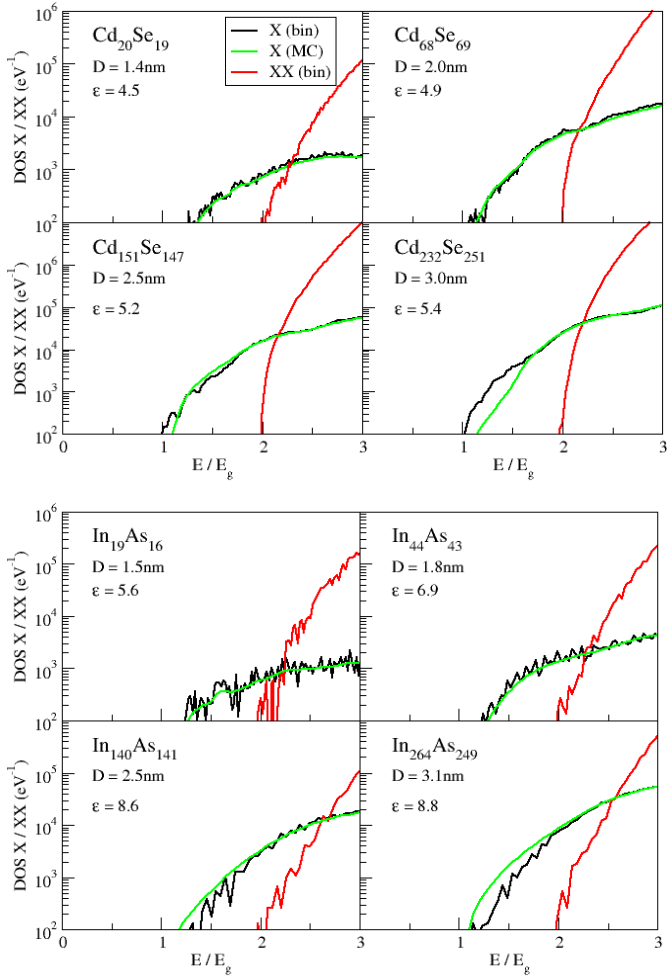

Figure 1: The density of single (DOSX) and bi- (DOSXX) excitons in various CdSe (upper panels) and InAs (lower panels) NCs.

The calculated DOSX and DOSXX are shown in Figure 1 for CdSe and InAs NCs for various sizes. The excitonic threshold occurs by definition at $E = E_g$. The two methods of calculating the DOSX agree well, indicating that the FD method is well converged, and all states are generated within the energy window up to $3E_g$. The bi-excitonic threshold is $2E_g$. For higher energies the DOSXX grows with energy at a considerably faster rate than the DOSX, overtaking it at scaled energies which only slightly depend on the size and composition of the NCS (between 2.3 and 2.5 $E_g$). The onset of CM in PbSe at around $2.2E_g$ has been attributed to this crossing [17]. However, it still remains an open question whether this crossing is indeed relevant for efficient CM. As we argue below, DOSXX is *not* the relevant density of states to consider because of the strict selection rules dictated by the exciton-biexciton coupling elements.

The process of CM involves the conversion of an exciton, of say spin up, $\left|S_{ia,\uparrow}\right\rangle = a^\dagger_{i\uparrow} a_{a\uparrow} \left|0\right\rangle$ to a biexciton $\left|B_{jbkc,\sigma''\sigma'''}\right\rangle = a^\dagger_{b\sigma''} a^\dagger_{c\sigma'''} a_{j\sigma'''} a_{k\sigma''} \left|0\right\rangle$ which is a state of two coexisting excitons. Here $\left|0\right\rangle$ is the ground-state determinant wave function where all hole states are occupied by electrons; $\hat{a}_{t\sigma}$ ($\hat{a}^\dagger_{t\sigma}$) are electron annihilation (creation) operators into the molecular orbital $\psi_t(\mathbf{r})$ with spin $\sigma$ (obtained from the pseudopotential calculation). In the following, we use the index convention that $i$, $j$ and $k$ designate hole orbitals; and $a$, $b$ and $c$ electron orbitals while $r$, $s$, $t$ and $u$ are general orbital indices. The rate of decay of a single exciton into biexcitons is give by Fermi's golden rule: $\Gamma_{ia\uparrow} = \frac{2\pi}{\hbar} \sum_{jkbc\sigma\sigma'} \left|W^{bkcj,\sigma\sigma'}_{ia}\right|^2 \delta\left(\left(\varepsilon_a - \varepsilon_i\right) - \left(\varepsilon_b - \varepsilon_k + \varepsilon_c - \varepsilon_j\right)\right)$ where $W^{bkcj,\sigma\sigma'}_{ia} = \left\langle S_{ia\uparrow}\right| \sum_{rsut\sigma''\sigma'''} \frac{1}{2} V_{rsut} a^\dagger_{r\sigma''} a^\dagger_{u\sigma'''} a_{t\sigma'''} a_{s\sigma''} \left|B_{jbkc,\sigma\sigma'}\right\rangle$ and $V_{rsut} = \iint d^3r\, d^3r' \left[\psi_r(\mathbf{r}) \psi_s(\mathbf{r}) \psi_u(\mathbf{r}') \psi_t(\mathbf{r}') / \varepsilon |\mathbf{r} - \mathbf{r}'|\right]$. Here $\varepsilon$ is the dielectric constant of the NC, estimated from [26] for CdSe and [21] for InAs. Deploying Fermionic commutation rules and energy conservation requirements it is possible to show the decay of exciton $S_{ia,\uparrow}$ to bi-exciton occurs in two types of fundamental processes. The decay of an electron (hole) in state $\psi_a$ ($\psi_i$) of energy $\varepsilon_a$ ($\varepsilon_i$) decays to a negative (positive) trion, composed of two electrons (holes) in states $\psi_b$ and $\psi_c$ ($\psi_j$ and $\psi_k$) and a hole (electron) in state $\psi_j$ ($\psi_b$). The trion must have the same energy as the electron (hole) so $\varepsilon_a = \varepsilon_b + \varepsilon_c - \varepsilon_j$ ($\varepsilon_i = \varepsilon_k + \varepsilon_j - \varepsilon_b$). The total decay rate is written as the sum of rates $\Gamma_{ia} = \Gamma^+_i + \Gamma^-_a$, given by Fermi's golden rule:

$$\begin{aligned}\Gamma^+_i &= \frac{4\pi}{\hbar} \sum_{jkb} \left|\left(2V_{jikb} - V_{kijb}\right)\right|^2 \delta\left(\varepsilon_i - \left(\varepsilon_k + \varepsilon_j - \varepsilon_b\right)\right) \\ \Gamma^-_a &= \frac{4\pi}{\hbar} \sum_{cbj} \left|\left(2V_{acjb} - V_{abjc}\right)\right|^2 \delta\left(\varepsilon_a - \left(\varepsilon_b + \varepsilon_c - \varepsilon_j\right)\right)\end{aligned} \quad (1)$$

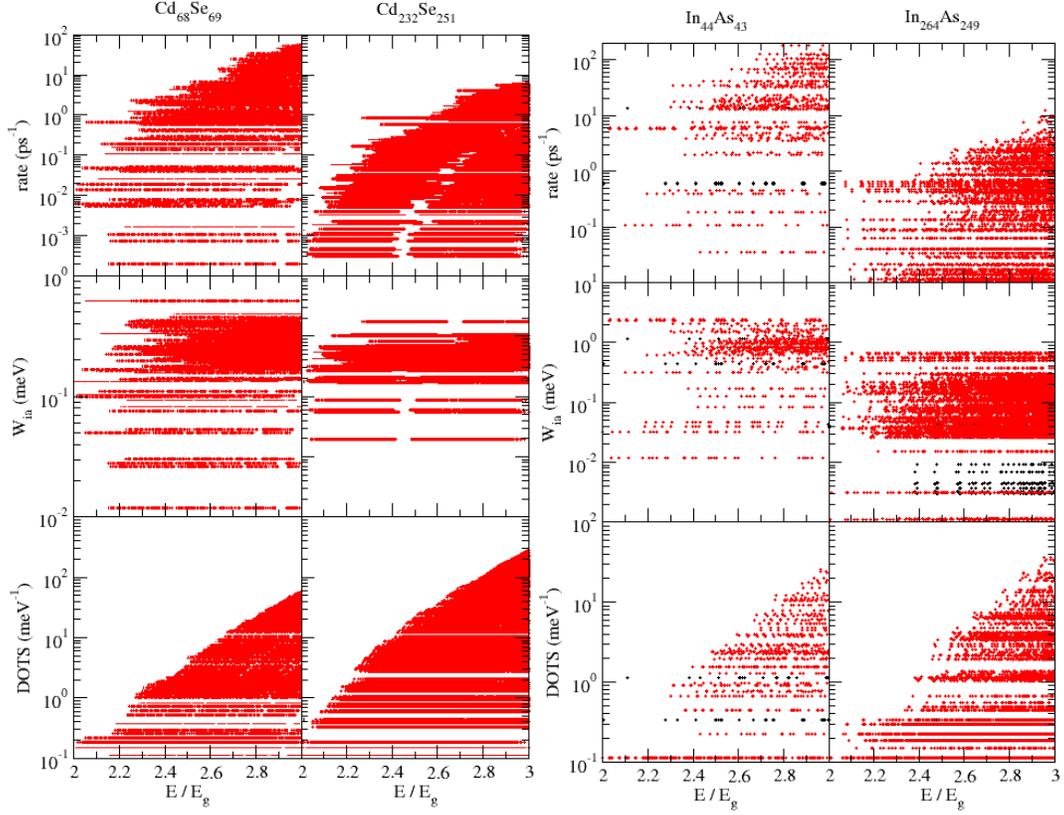

Figure 2: The rate of exciton-biexciton transition, Coulomb coupling and TDOS for each exciton in the energy range of $2-3\,E_g$ for two NCs of CdSe and InAs.

In Figure 2, we show results for CdSe (left panels) and InAs (right panels) at two sizes of NCs for exciton energies in the $2-3E_g$ range. Each point in the figure represents an exciton $\left|S_{is}\right\rangle$ of the with a scaled energy $\left(\varepsilon_a - \varepsilon_i\right)/E_g$. The lowest panel depicts the density of trion states (DOTS): positive trions are black points with DOTS $\rho_{ia}^+ = \sum_{jkb}\delta\left(\varepsilon_i + \varepsilon_b - \varepsilon_k - \varepsilon_j\right)$, and negative trions are red points with DOTS $\rho_{ia}^- = \sum_{cbj}\delta\left(\varepsilon_a + \varepsilon_j - \varepsilon_b - \varepsilon_c\right)$. It is seen that the number of black points is much smaller than the number of red points. Similar to the situation in the bulk, the low mass particle takes most of the exciton energy, which for CdSe and InAs is the electron. Thus there are many more negative trions in resonance with the electron than positive trions in resonance with the hole. In the bulk the lighter particle takes most of the kinetic energy because both must have equal momentum. In NCs with strong confinement this result is due to the fact that the DOS near the band edge grows with the particle mass, so heavier (lighter) particles have a large (small) density of states near the band edge.

The upper panels in Figure 2 shows decay rate for each exciton (see Eq. (1)): $\Gamma_{ia}^+$ via a positive trion or $\Gamma_{ia}^-$ via a negative trion. The effective Coulomb matrix element, defined as $W_{ia}^\pm = \sqrt{\hbar\,\Gamma_{ia}^\pm/\left(2\pi\rho_{ia}^\pm\right)}$, is shown in the center panels.

There are several important conclusions drawn from the results shown in Figure 2:

(a) In CdSe the DOTS increases at a given scaled energy as the size of the NC grows. In InAs this behavior is much weaker. At a given energy the DOTS of a NC increases with size, however, at a *scaled* energy, since $E_g$ *decreases* with size, such size dependence is weaker. In InAs the strong confinement makes $E_g$ highly sensitive to size causing a reduced sensitivity of the DOTS as a function of the scaled energy. Overall there are fewer trion states for InAs compared to CdSe because $E_g$ is smaller in the former NC thus the absolute energy probed is lower.

(b) The effective coupling $W_{ia}^\pm$ for a given exciton is nearly energy independent, approximately equals to 0.1-1 meV depending on the size of the NC, with a spread that decreases with NC size and spans 1-2 orders of magnitude. NCs of smaller diameter $D$ exhibit larger coupling elements. However, the coupling is not proportional to $D^{-1}$ as expected when the states scale linearly with $D$.

(c) The rate of exciton-biexciton transformation at a given exciton energy spans 4-6 orders of magnitude, depending on the specific exciton ($i$ and $a$). Thus, conclusions re-garding the CM process require the calculation of the rate for *all* excitons in a given energy, and may not be drawn from a limited arbitrary set of excitons assumed in previous studies by others. We find that due to quantum confinement, the smaller NCs span a larger range of rates. In addition, smaller NCs have smaller DOTS but larger $W_{ia}^\pm$. The net effect of combining the two quantities into the rate results in a larger rate for smaller NCs at a given scaled energy.

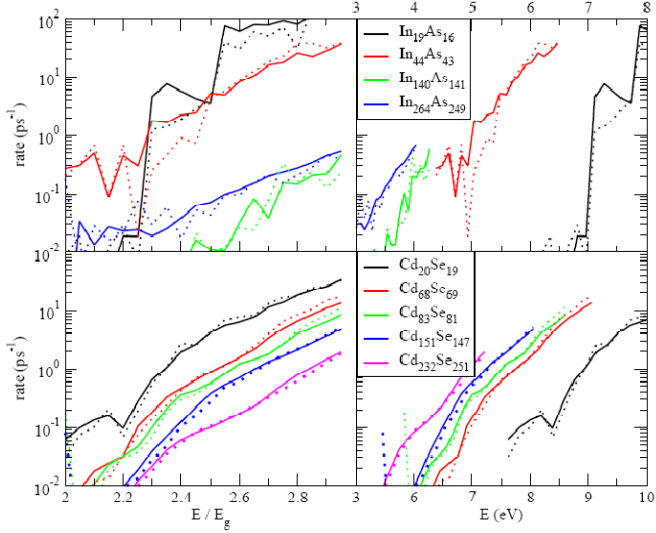

Figure 3: The average rate of exciton to biexciton transition for various CdSe NCs (lower panels) and InAs NCs (upper panels) as a function of scaled exciton energy $E/E_g$ (left panels) and absolute exciton energy $E$ (right panels). Solid and dashed lines represent two methods of averaging as discussed in the text.

In Figure 3 we show the average rate of exciton to biexciton transition for various CdSe and InAs NCs. Two averaging schemes are used yielding similar results. One is a straightforward arithmetic average (solid lines) and the other is an oscillator strength weighted average (dotted line). As discussed above, the transition rate decreases with NC size $D$ at a given *scaled energy*, but it increases with $D$ at a given *absolute energy*. This can be traced to a simple different stretching of the energy axes, as seen when comparing the left and right panels in Figure 3.

There are two main factors that influence the dependence of CM rate on the size of the NC, the DOTS and the effective coupling $W_{ia}^\pm$. As discussed above, the DOTS increases with energy and mildly with $D$ while the effective coupling decreases strongly with $D$ (due to the decreasing Coulomb matrix element and the increasing dielectric constant) and is nearly exciton-energy independent. The CM rate, which is proportional to the product of the two, inherits its dependence on the energy from the DOTS while its dependence on $D$ from $W_{ia}^\pm$.

CM should be observed when the rate of exciton-biexciton transition is faster than the rate of exciton decay by other channels, i.e., faster than $\approx 1 ps^{-1}$. This condition sets a threshold for efficient CM, which is $E > 2.3 E_g$ for the smallest NCs ($D \approx 1.5 nm$) and $E > 3 E_g$ for the largest NCs considered here ($D \approx 3 nm$). Therefore, in the energy range of 2-3 $E_g$ CM will only be efficient for small NCs consistent with the *ab inito* calculations of ref. [19], but as the size increases CM at this energy range is unlikely to occur. The latter conclusion is consistent with known results for bulk. They are also in agreement with recent experimental results on CdSe [13] and InAs [14] for NCs of $D \geq 5 nm$. The measurements on CdSe NCs with $D = 3.2 nm$ [6] is a border case, and we predict that CM may occur for this system depending on the value of $\gamma_1$. However, we show that at $E = 2.5 E_g$ CM is of very low efficiency and is possible only at $E \geq 3 E_g$. Our results disagree with one experiment, namely the positive CM in InAs [11] at $D = 4.3 nm$. We argue that the onset of CM should be observable only for energies larger than $3 E_g$ but not at $2 E_g$ where our predictions for the exciton-biexciton transition rate is $0.01 ps^{-1}$. Other materials require explicit calculation of the CM rate however, we anticipate that a similar picture will emerge.

In summary, we have carried out detailed calculations of the exciton-biexciton transition rate using Fermi's golden rule for CdSe and InAs NCs at different sizes and in the energy range of $2 - 3 E_g$. We use the highly reliable semiempirical atomistic electronic structure method and introduce Coulomb coupling between excitons and biexcitons in a consistent way via a perturbation theory. We do not find evidence that the CM is correlated with the crossover of DOSX and DOSXX, since the relevant density of states entering Fermi's golden rule is the DOTS. We predict that there is a wide spread of rates (several orders of magnitude) for different excitons at a given energy dominated by decay to negative trions. The average rate is strongly size and energy dependent. For CdSe and InAs NCs with diameter larger than 3 nm we argue that CM below $3 E_g$ is of very efficiency, but at higher energies or smaller NCs, CM can become efficient.

This research was supported by the Converging Technologies Program of The Israel Science Foundation (grant number 1704/07) and The Israel Science Foundation (grant number 962/06).